\documentclass{svmult}
\bibstyle{spphys}

\usepackage{mathptmx}       
\usepackage{helvet}         
\usepackage{courier}        
\usepackage{type1cm}        
\usepackage{makeidx}         
\usepackage{graphicx}        
\usepackage{multicol}        
\usepackage[bottom]{footmisc}
\usepackage{epstopdf}

\usepackage{subfigure}

\makeindex

\begin{document}

\title*{An Implementation of
Voice over the Cognitive Packet Network}

\author{Lan Wang and Erol Gelenbe}

\institute{Imperial College London
Department of Electrical and Electronic Engineering
Intelligent Systems and Networks Group, \email{{lan.wang12,e.gelenbe}@imperial.ac.uk}}
%
%
\maketitle

\abstract{Voice over IP (VOIP) has strict Quality of Service (QoS) constraints and requires real-time packet delivery, which poses a major challenge to IP networks. The Cognitive Packet Network (CPN) has been designed as a QoS-driven protocol that addresses user-oriented QoS demands by adaptively routing packets based on online sensing and measurement. This paper presents our design, implementation and evaluation of a ``Voice over CPN'' system where an extension of the CPN routing algorithm has been developed to support the needs of voice packet delivery in the presence of other background traffic flows with the same or different QoS requirements.}

\keywords{Cognitive Packet Network, Voice over CPN, Quality of Service, Experimental Implementation}

 \section{Introduction}
\label{INTRO}

In the current ``All IP'' era, IP networks are required to guarantee Quality of Service (QoS) for a vast variety of communication services and users~\cite{Toral01}, especially for real-time voice services which have stringent QoS constraints. In the past decade, many QoS approaches such as  IntServ\&RSVP, DiffServ and MPLS were proposed yet with limited effect. While networks must be monitored and measured \cite{FCC,MeasurementLab} for performance in terms of loss \cite{Loss}, delay, packet desequencing \cite{Christophe01}, for topology purposes \cite{Zhang05} and reliability \cite{Failures}, mathematical models are also used \cite{Acta79} together with measurements and simulations.

The Cognitive Packet Network (CPN) has been designed as a QoS-driven protocol that addresses user-defined QoS demands by routing packets in the manner that adapts to the varying network conditions based on online sensing and measurement~\cite{Nagoya1,Reif,Autonomic04}. It was developed under the open-source platform Linux, incorporated into IP protocols, and running as a loadable kernel module at any linux machine, which made it feasible to construct a large and measurable CPN testbed carrying many network services. It has also been used for organised exit paths for emergency management operations \cite{HSI,DBES}.

In CPN, QoS requirements specified by users, such as $Delay$, $Loss$, $Energy$ \cite{gelenbe-mahmoodi-Energy,Power2}, or a combination of the above, are incorporated in the ``goal'' function which is used for the CPN routing algorithm. Three types of packets are used by CPN: smart packets (SPs), dumb packets (DPs) and acknowledgments (ACKs). SPs explore the route for DPs and collect measurements; DPs carry payload and also conduct measurements; ACKs bring back source routing information for the DPs. SPs discover the route using random neural network (RNN)-based reinforcement learning (RL)~\cite{RNN-Learning,CompNet01,SAN04} which resides in each node. Each RNN in a node corresponds to a QoS class and destination pair \cite{CACM} with each neuron representing a outgoing link from the node. The arrival of an SP for a specific QoS class at a node triggers the execution of the RNN algorithm with the weights updated by Reinforcement Learning (RL) using QoS goal-based measurements, whereby routing decision are based on selecting the output link corresponding to the most excited neuron~\cite{Mascots02,sp_acc05}.



\begin{figure}[ht]
		\centering
			\includegraphics[scale=0.35]{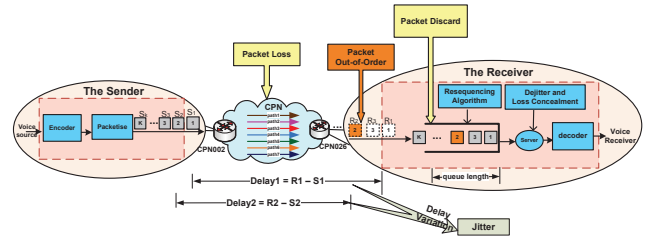}
		\centering
		\caption{VOCPN System Structure ~\cite{Baccelli01}}
		\label{fig:VoIPStructure}
		\end{figure}
		
 At the sender in a VOIP application installed at a CPN node, the original analog voice signals are sampled, encoded and then packetised into IP packets by adding RTP header, UDP header, and IP header. These packets travel across the IP network employing the CPN protocol, where IP-CPN conversion is performed at the source node by encapsulating IP packets into CPN packets. Previous research has reported the ability of CPN to alleviate delay, jitter, packet desequencing and packet loss for real-time traffic by smartly selecting the path that provides the best possible QoS required by a user (or an application)~\cite{phd-gellman, Mascots02} and thus the packets in a voice traffic flow may traverse different paths over the CPN network. At the receiver, packets are queued in a buffer to reduce jitter while reordering algorithm and loss concealment techniques are applied before the recovery of the original voice signals. Packets that arrive later than the required playback time or those that provoke buffer overflow, are discarded, contributing to the end-to-end packet loss.

This paper presents an extension of CPN structured as in Figure~\ref{fig:VoIPStructure}, to support the QoS of voice packet delivery in the presence of other background traffic flows with the same or different QoS requirements. The resulting system was evaluated in order to select the QoS goal that provides better performance for all network load conditions. Furthermore, we  study the correlation of voice packet end-to-end loss with path switching induced by the adaptive scheme that is inherent to CPN.

\section{Voice over CPN Supporting Multiple QoS Classes} \label{VOCPN}

This section presents the incorporation of ``Jitter'' minimisation into the goal function used by CPN in order to match the needs of voice delivery, as well as the extension of CPN that supports multiple QoS classes for multiple traffic flows simultaneously. In RFC3393 and RFC5481 , ``Packet Delay Variation'' is used to refer to ``Jitter''. One of the specific formulations of delay variation implemented in the industry is called Instantaneous packet delay variation (IPDV)
which refers to the difference in packet delay between successive packets, where the reference is the previous packet in the stream's sending sequence so that the reference changes for each packet in the stream.

The measurement of IPDV for packets consecutively numbered $i = 1,2,3,...$ is as follows. If $S_i$ denotes the departure time of the ith packet from the source node, and $R_i$ denotes the arrival time of the ith packet at the destination node, then the one-way delay of the ith packet $D_i=R_i-S_i$, and IPDV is
\begin{equation}
\label{eq:IPDV_T}
	IPDV_i= |D_i-D_{i-1}|= |(R_i-S_i)-(R_{i-1}-S_{i-1})|
\end{equation}

To fulfill the QoS goal of minimising jitter, online measurement collects the jitter experienced by each dumb packet. Since in CPN each DP carries the time stamp of its arrival instant at each node along its path, so when a DP
say $DP_i$ arrives at the destination, an ACK is generated with the arrival time-stamp provided by the DP.
As $ACK_i$ heads back along the inverse path of the DP, at each node the forward delay $Delay_i$ is estimated from this node to the destination by taking the difference between the current arrival time of $ACK_i$ at the node and the time at which the $DP_i$ reached the same node~\cite{Mascots02}, divided by two. This quantity is deposited in the mailbox at the node. The instantaneous packet delay variation is computed as the difference between the value of $Delay_i$ and $Delay_{i-1}$ of the previous packet in the same traffic flow as defined in (\ref{eq:IPDV_T}),
and jitter is approximated by the smoothed exponential average of IPDV with factor a smoothing factor $0.5$:
\begin{equation}
\label{eq:JITTER_AVG}
	\overline{J_i} = \frac{J_{i-1}}{2} + \frac{J_i}{2}
	\end{equation}
Then, the updated value of jitter is deposited in the node's mailbox. When a subsequent SP for the QoS class of Jitter and the same destination enters the node, it uses the value from the mailbox to compute the reward $Reward_i$ and in turn update the weights of the corresponding RNN which is then executed to decide the outgoing link~\cite{Mascots02}.
\begin{equation}
\label{eq:REWARD}
Reward_i = \frac{1}{\overline{J_i}+\epsilon}
\end{equation}

\noindent where $\epsilon$ is used to ensure the denominator is non-zero.


We enable CPN to support multiple QoS classes simultaneously, for multiple flows that originate at any node and each flow is routed based on its specific QoS criteria, the following steps are needed: 1) The traffic differentiation is conducted relying on source MAC (or IP), destination MAC (or IP) and the TCP/UDP port of the applications. For instance, the VOIP application ``Linphone'' originates voice packets with its dedicated SIP port (5060) and audio port (7080) residing in the fields of the UDP header.
2) The QoS class definition is based on the QoS requirements of different users or applications, which is configurable and loaded into the memory while CPN is being initiated.
3) CPN treats each traffic flow according to its QoS class using multiple RNNs at each node, where each RNN corresponds to a QoS class and a source-destination pair.

\section{Path Switching, Packet Reordering and Loss}
\label{PacketLossandPathSwitching}

CPN adaptively selects the path that provides best possible QoS requested for traffic transmission, leading to
possible path switches. Traffic may suffer packet desequencing and loss if paths switch excessively. Accordingly, we are interested in examining the correlation between undesirable effects such as packet desequencing and end-to-end loss, and path switching. In the following sections, we described methods to carry out measurements and statistics for the three metrics.

For a given flow in CPN, the routing information explored by SPs is encapsulated into the CPN header for each DP as it originates from the source node, whereby the path used by each DP can be detected. The metric we are interested in is the ``Path Switching Ratio'', which is defined as the number of path switches ($Q_{path}$) in a given flow during the time interval being considered being divided by the total number of packets forwarded ($N$), as well as the ``Path Switching Rate'' ($Rate_{path}$) defined as $Q_{path}$ being divided by the time interval ($T$).

	
	Packet reordering is an important metric for voice because packets have to be played back sequentially at the receiver in the same order that they have been sent. Thus, packets arriving earlier than their predecessors have to be buffered for reordering. We measure it according to the recommendation from~\cite{RFC4737}, which is based on the monotonic ordering of sequence numbers with a constant increment (denoted by $Seq_{inc}$). Specifically, in CPN packets are identified successively in the sending sequence with increment of ``1''. To detect packet reordering, at the receiver we reproduce the sender's identifier function where the variable $NextExp$ is used to represent the Next Expected Identifier which is incremented by $Seq_{inc}$ once the in-order packet arrives. Given $S$ is the identifier of the current arrival, if $S < NextExp$, the packet is reordered, else update $NextExp \leftarrow S+Seq_{inc}$.
	
	 To quantify the degree of desequencing, we also defined the ``Packet Reordering Ratio/Rate'' and the ``Packet Reordering Density'' denoted by $Density_r$, so that we may differentiate between isolated and bursty packet reordering as well as to measure the degree of burstiness of packet reordering, which may affect
	the packet drop rate of the resequencing buffer at the receiver. $Density_r$ is calculated as:

\begin{equation}
\label{eq:density}
Density_r += \left\{
  \begin{array}{l l}
     Cout_r^2  & for~bursty~packet~reordering\\
     Cout_r  &  for~isolated~packet~reordering .
  \end{array} \right.
\end{equation}
\noindent where $Cout_r$ is the number of successively reordered packets; it resets to zero when the in-order packets arrive and is incremented when reordering occurs.

The recommendation in \cite{RFC3611} states that packet loss should be reported ``seperately on packets lost in a network, and those that have been received but then discarded by the jitter buffer'' at the receiver for real-time packet delivery, because both have an equal effect on the quality of voice services, which is also denoted as packet end-to-end loss.

	Packet loss within the network is detected for a packet that is sent out but not received by its destination node based on the matching of the packet identifier, the source and destination IP address. The resequencing buffer at the receiver is necessarily of finite length so that packets arriving to a buffer that is full will be discarded, and packets will have to be forwarded after a given time-out even when their predecessors have not arrived in order to avoid excessive time gaps with their predecessors that have already been played back. Thus, packets that arrive later than the expected playback time will also be discarded. As we cannot get directly access to the run-time version of the VOIP application, we had to simulate the operation of a jitter buffer which employs resequecing so as to study packet discards and the buffer queue length, and their correlation with packet reordering and packet loss. We also defined the ``Packet (end-to-end) Loss Ratio/Rate/Density'' consistent with (\ref{eq:density}).

		
\section{Experimental Results}
\label{ExperimentalResults}

\begin{figure}[ht]
		\centering
			\includegraphics[scale=0.3, clip=true, trim=0 400 0 60, width=6cm,height=3.5cm]{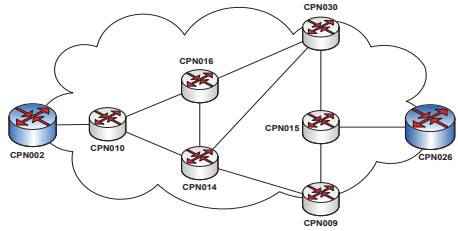}
		\caption{CPN testbed network topology in the experiment}
		\label{fig:TestbedCPNTopology}
		\end{figure}

Our experiments were carried out on a wired test-bed network consisting of 8 nodes with the topology shown in Figure~\ref{fig:TestbedCPNTopology}, whereby multiple paths are available for packets delivery between arbitrary source-destination pair. CPN was implemented as a loadable kernel module~\cite{CACM} running under linux 2.6.32 at each node. Adjacent nodes are connected with 100Mbps Ethernet links.

We installed ``Linphone'', a VOIP phone, at each node in the network testbed to generate actual voice traffic while UDP traffic generation applications are running to introduce background traffic flows with a range of data rates to vary load conditions. As human listeners of voice are sensitive to the time-based QoS metrics ``delay'' and ``delay variation'', our experiments were conducted with voice traffic and one of the two QoS requirements, in the simultaneous presence of several background traffic flows with the same or the other QoS goal for the duration of ten minutes. We repeated each experiment with data rates of 1M, 2M, 3.2M, 6.4M, 10M, 15M, 20M, 25M, 30M bps and packet size of 1024 bytes for background traffic. Measurements of the voice traffic flow were gathered between CPN002 and CPN026 as this source-destination pair has the most intermediate nodes.
	\begin{figure}[ht]
	
		\centering
			\subfigure{\includegraphics[scale=0.25,width=5.5cm,height=3cm]{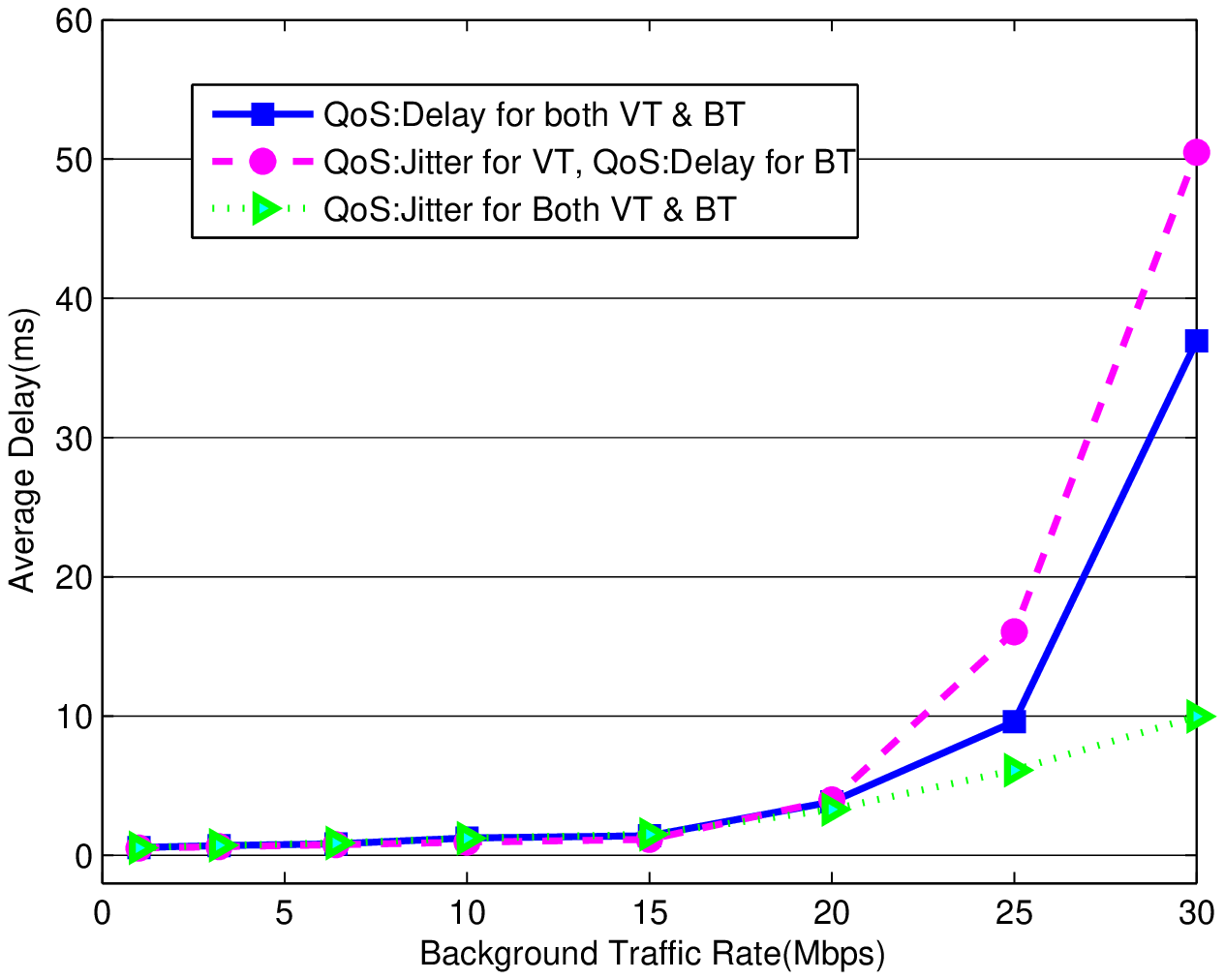}}
			\subfigure{\includegraphics[scale=0.25,width=5.5cm,height=3cm]{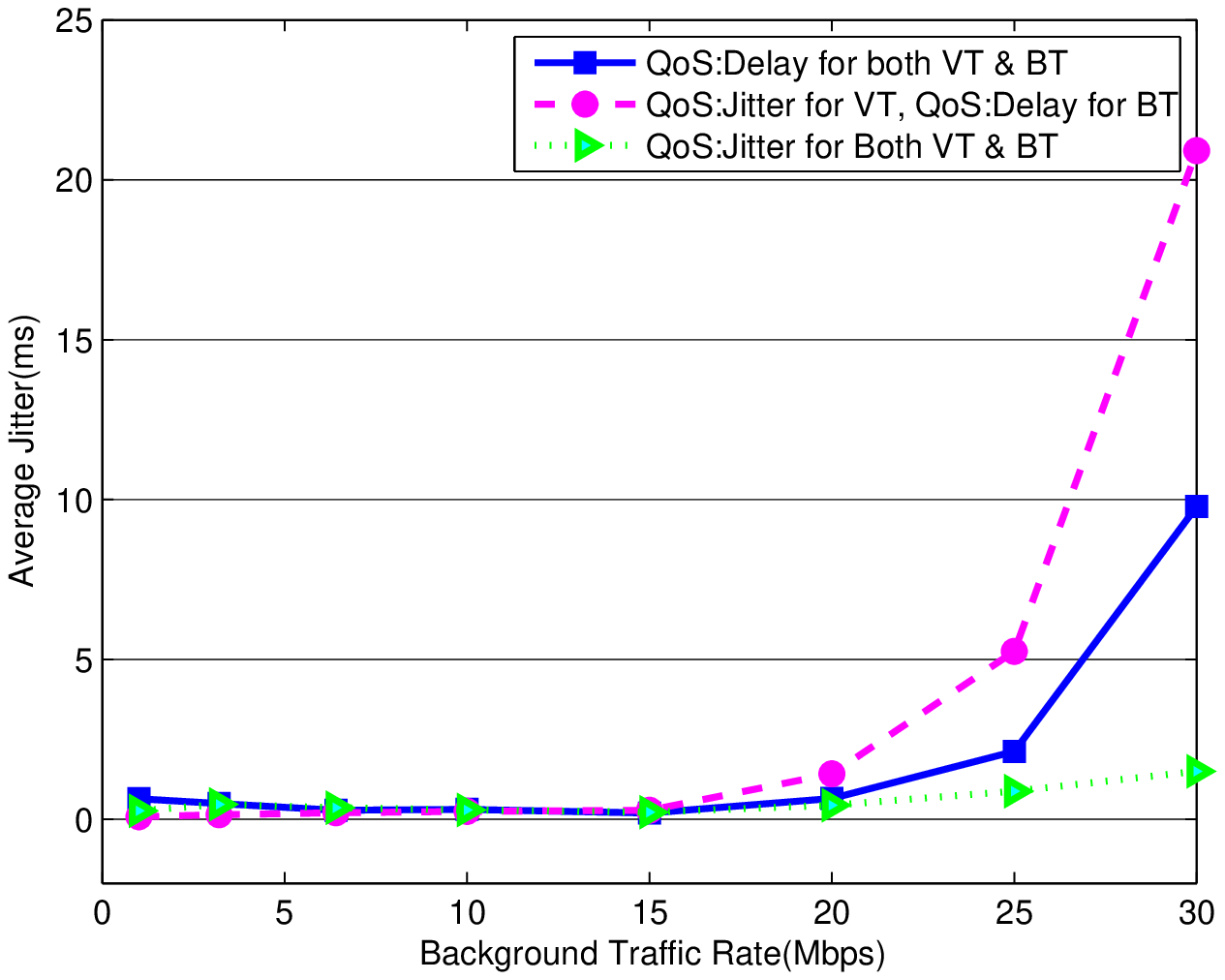}}
		\caption{The performance for Voice Traffic under varied background traffic conditions}
		\label{Fig:performance}
		\end{figure}

\begin{figure}[ht]
		\centering
			\subfigure{\includegraphics[scale=0.25, width=5.5cm,height=3cm]{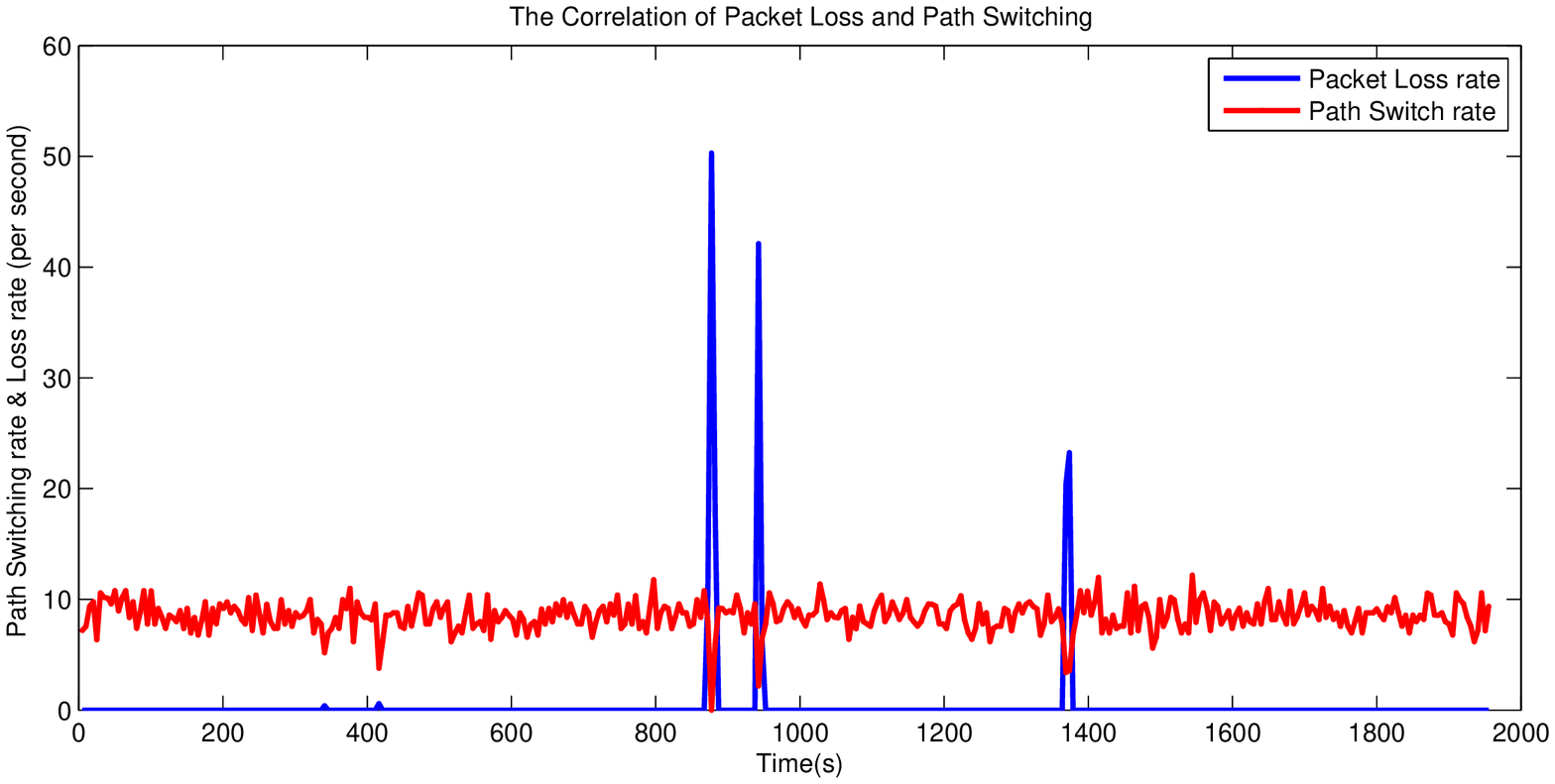}}
			\subfigure{\includegraphics[scale=0.25, width=5.5cm,height=3cm]{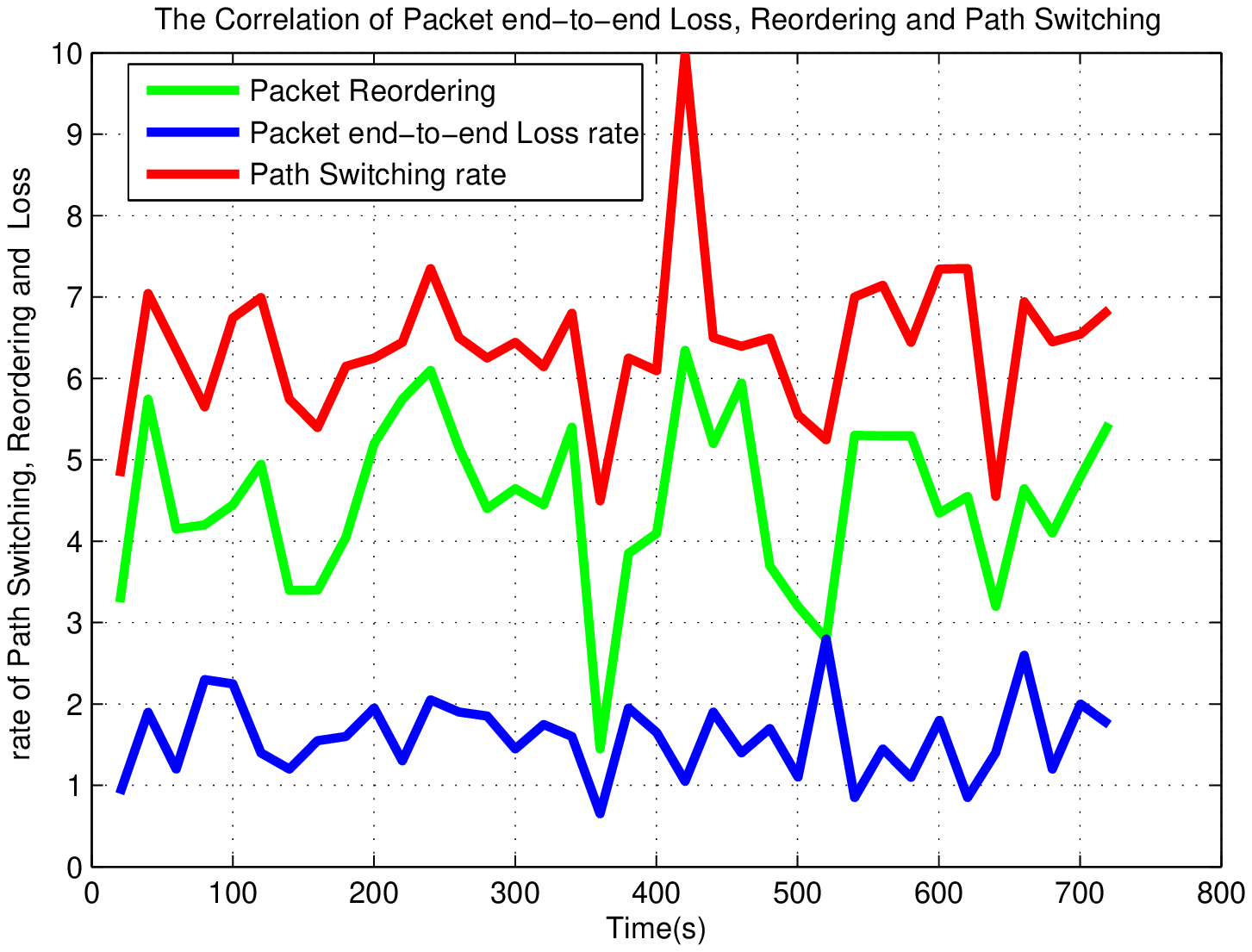}}
		\caption{The correlation of Packet Loss and Path Switching under medium(L) and heavy(R) traffic condition}
		\label{Fig:losspswitching}
		\end{figure}

	The measurements shown in Figure~\ref{Fig:performance} indicate that when we use Jitter Minimisation as the QoS goal both for the voice itself and the background traffic, jitter appears to be indeed minimised but also delay and traffic loss are reduced for the voice traffic, because path switching is also reduced,  alleviating route oscillations at heavy traffic loads.  From observations for the voice traffic flow between CPN002 and CPN026 during the test interval considered while increasing the rate of the background traffic gradually, as shown in Figure~\ref{Fig:losspswitching} (Left) as the rate of background traffic reached 20Mbps,
	it is seen that in the
intervals (800s-900s, 900s-1000s, 1300s-1400s), bursty packet loss occurred when path switching rates were low. This seems to arise from the fact that when a given path used by voice traffic satisfies the QoS criterion for a long time, and the path switching rate is close  to ``0'', this path attracts other traffic, becoming saturated with packet loss ratio reaching ``1''. Subsequently, the performance degradation is detected by the SPs and they move the traffic to less loaded paths.

As the rate of the background traffic increased to 30Mbps, We can see from Figure~\ref{Fig:losspswitching} (Right) that the occurrence of packet desequencing was frequent and varied proportionally with packet path switching, which demonstrates that packet reordering is mainly due to path switching in CPN. It is not easy to observe the correlation of packet path switching and packet loss from the figure. It is possibly because under heavy traffic conditions, packet loss is not only due to link saturated, route oscillation induced by heavy traffic loads also leads to the occurrence of loss. We can also found that an adequate path switching rate is beneficial to loss reduction, but if path switching rate is increased excessively, it is converted to route oscillation which also lead to packet loss.
		To evaluate the packet drops at the receiver, we also used the voice packets received at CPN026 with the background traffic at rate of 30Mbps as the input data to the simulation. It was found that the bursty packet loss and reordering that can be provoked by path switching within a network, i.e. the successive occurrence of packet loss and reordering, will induce delays for other packets in the output resequencing buffers of the VOIP codec, which in turn can provoke buffer overflow and further losses.




\bibliographystyle{spphys}
\bibliography{paperref}


%



\end{document}